\documentclass[prl,superscriptaddress,amsmath,amsfonts,twocolumn]{revtex4}
\usepackage{amsfonts}
\usepackage{times}
\usepackage{color}
\usepackage{epsfig}

\begin{document}
\title{Dynamical mechanism of anticipating synchronization in excitable systems}
\author{Marzena Ciszak}
\affiliation{Departament de F\'{\i}sica, Universitat de les Illes
Balears}
\author{Francesco Marino}
\affiliation{Instituto Mediterr\'aneo de Estudios
Avanzados (IMEDEA), CSIC-UIB, Ed. Mateu Orfila, Campus UIB, 07122
Palma de Mallorca, Spain}
\author{Ra\'ul Toral}
\author{Salvador Balle}
\affiliation{Departament de F\'{\i}sica, Universitat de les Illes
Balears}
\affiliation{Instituto Mediterr\'aneo de Estudios
Avanzados (IMEDEA), CSIC-UIB, Ed. Mateu Orfila, Campus UIB, 07122
Palma de Mallorca, Spain}
\homepage{http://www.imedea.uib.es}

\date{\today}

\begin{abstract}
We analyze the phenomenon of anticipating synchronization of two excitable
systems with unidirectional delayed coupling which are subject to the same
external forcing. We demonstrate for different paradigms of excitable system
that, due to the coupling, the excitability threshold for the slave system is
always lower than that for the master. As a consequence the two systems respond 
to a common external forcing with different response times. This allows to explain in a simple way the
mechanism behind the phenomenon of anticipating synchronization.
\end{abstract}

\pacs{}

\maketitle


The synchronization of nonlinear dynamical systems is
a phenomenon common to many fields of science ranging from biology
to physics\cite{PRK01}, and it has been an
active research subject since the work by Huygens in $1665$. Recently, the synchronization of chaotic
systems in a unidirectional coupling configuration has attracted a
great interest due to its potential applications to secure
communication systems \cite{PCJ97}. Particular attention has been
payed to the so-called {\sl anticipating synchronization} regime,
an idea first proposed by Voss in \cite{voss1}. He showed that, in
some parameter regions, two identical chaotic systems can be
synchronized by unidirectional delayed coupling in such a manner
that the "slave" (the system with coupling)
anticipates the "master" (the one without coupling). More
specifically, the coupling scheme proposed in 
\cite{voss1} for the dynamics of the master, $x(t)$, and slave, $y(t)$ is the
following:
\begin{eqnarray}
\dot x & =  & F(x)\\
\dot y & =  & F(y)+K(x-y^{\tau})
\end{eqnarray}
where $y^{\tau} \equiv y(t - \tau)$. 
For appropriate values of the delay time $\tau$ and 
coupling strength $K$, the basic result is that $y(t)\approx x(t+\tau)$, 
i.e. the slave ``anticipates'' by an amount $\tau$ the output of the master.

This regime and its stability has been theoretically studied in several
systems, from the simplest ones described by linear differential equations and
maps where the mathematical details can be fully worked out\cite{hern,calvo},
to the more complicated ones such as semiconductor lasers \cite{cris}
operating in the chaotic regime. Experimental evidence of anticipating
synchronization has been shown in Chua circuits \cite{voss4} and in
semiconductor lasers with optical feedback \cite{liu}.

This same phenomenon has recently been shown to occur also when
the dynamics, instead of chaotic, is excitable. In refs. \cite{marzena} the effects of unidirectional delayed coupling between
two identical excitable systems was studied for both the
FitzHugh-Nagumo \cite{fn} and Hodgkin-Huxley \cite{hh} models. It
was shown that, when both systems are excited by the same noise, and for a certain range of coupling parameters, the
randomly distributed pulses of the master are preceded by
those of the slave. This allows for
predicting the occurrence of excitable pulses in the master.
Since many biological systems (as neurons and heart cells) exhibit
excitable behavior and they often operate in feedback regime in a
noisy environment, the study of the delayed coupling effects in a
presence of noise is certainly of wide concern.

The anticipating synchronization regime 
has been often described as a rather counterintuitive phenomenon because of
the possibility of the slave system anticipating the unpredictable evolution of
the master \cite{voss1,calvo,voss4}. The aim of this paper is to provide a 
simple clear physical mechanism for this regime in delayed coupled excitable
systems, showing that the anticipation of the slave is due to a
reduction of its excitability threshold induced by the delayed
coupling term. As a consequence, the master and the slave respond
to a common external forcing with different response times. The proposed
dynamical picture allows us to explain all the general features of the 
phenomenon as well as to determine in a natural way the maximum permitted
anticipation time.
The results are sustained by numerical integration of the dynamical equations as
well as by simple analytical calculations.


A dynamical system commonly used to study excitable behavior is
Adler's equation, \cite{egmind}
\begin{equation}
\dot{x} = \mu - \cos x \; ,
\end{equation}
where $x$ is an angular variable (modulo $2\pi$) and $\mu$ the control parameter. For
$\vert \mu \vert < 1$, there are two fixed points at $x_{\pm} = \pm \arccos \mu$, one being
a stable focus ($x_-$) and the other ($x_+$) an unstable saddle point. If $|\mu| > 1$,
there are no fixed points, and the flow consists in an oscillation of the variable $x$.
This limit cycle develops through an Andronov bifurcation at $\mu_c = \pm 1$
\cite{coullet,andronov}, where the two fixed points collide and annihilate.
For $|\mu| < 1$, the system displays excitable behavior: if we kick the system out of its stable state
with a large enough perturbation, the trajectory  will return to the initial state (modulo
$2 \pi$) through an orbit that closely follows the heteroclinic connection of
the saddle and the node. During this orbit, the system is barely sensitive to external
perturbations of moderate amplitude.

\begin{figure}[h]
\begin{center}
\epsfig{file=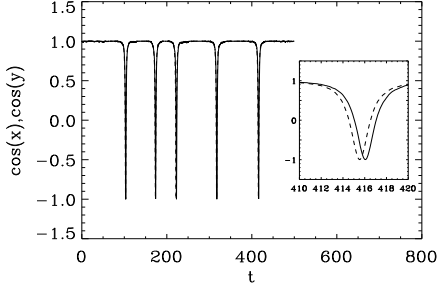,width=6cm,height=4cm}
\end{center}
\caption{\label{fig1} Time series of the master system $x$ (solid line) and
slave system $y$ (dashed line) subjected to white Gaussian noise of zero mean and correlations $\langle(\xi(t)\xi(t')\rangle=D\delta(t-t')$, obtained by numerical
simulation of eqs. (\ref{1},\ref{2}). Other parameters are: $\mu = 0.95$, $K=0.01$
$\tau = 1$. The noise intensity is $D = 0.017$.}
\end{figure}
In order to study anticipating synchronization, we consider two identical
Adler's systems with delayed unidirectional coupling under the effect of an
external perturbation $I(t)$ acting simultaneously on both systems,
\begin{eqnarray}\label{1}
\dot{x} & = & \mu - \cos(x) + I(t)\\\label{2}
\dot{y} & = & \mu - \cos(y) + K (x - y^{\tau}) + I(t)
\end{eqnarray}

When $I(t)=\xi(t)$ is zero-mean Gaussian noise, anticipating
synchronization occurs as shown in Fig. \ref{fig1}, where we plot
the master and slave outputs for a particular value of $K$ and
$\tau$. Note that the slave system anticipates the firing of a
pulse in the master by a time interval approximately equal to
$\tau$. If we increase the coupling constant $K$ or the delay time
$\tau$ beyond some values, anticipating synchronization is
degraded, i.e., the slave system can emit pulses which do not have
a corresponding pulse in the master's output,
although the reverse case never occurs. Upon further increasing $K$ or
$\tau$, the anticipation phenomenon disappears. The results are analogous
 to those obtained in
\cite{marzena} for the FitzHugh-Nagumo model.

\begin{figure}[h]
\begin{center}
\hspace{-0.5cm}\makebox{\epsfig{file=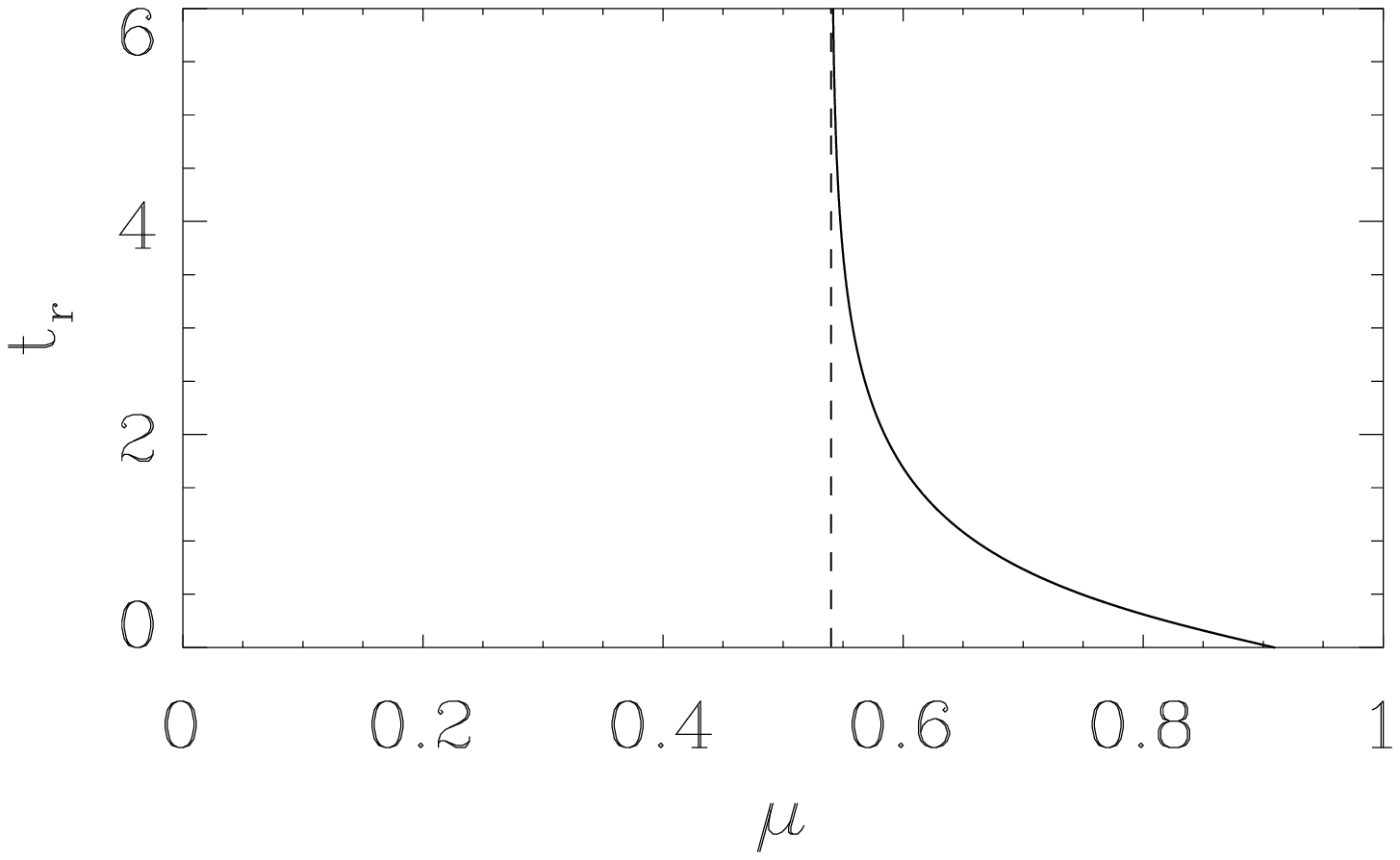,width=4.5cm,height=3.5cm}
\epsfig{file=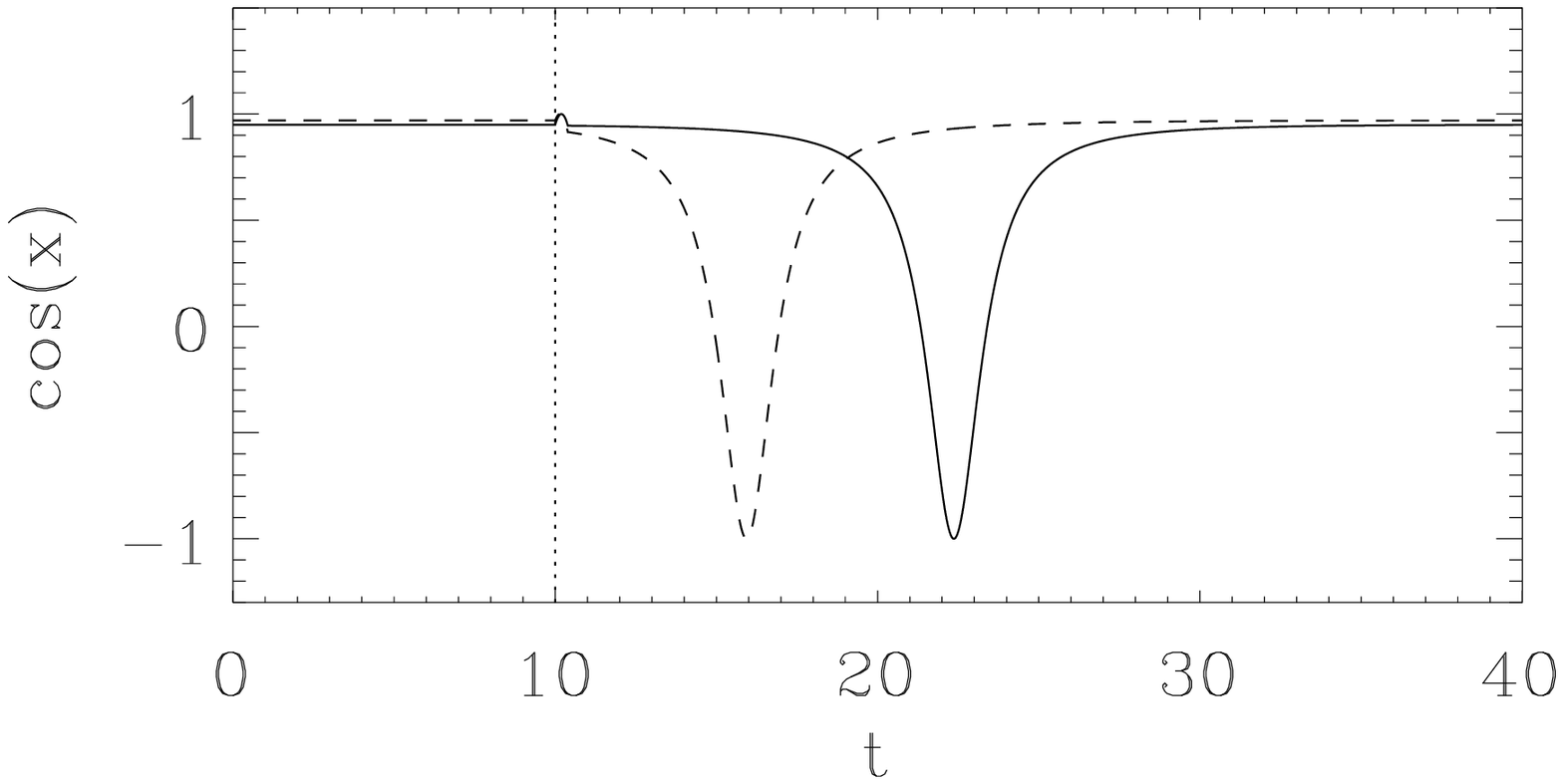,width=4.5cm,height=3.5cm
}}\end{center}
\caption{\label{fig2} Left panel: Response time $t_r$ versus $\mu$ for the Adler system $x$
perturbed by $p_0\delta (t)$ with $p_0=2$ from Eq. \ref{eq:tr}. Right panel: time series for $x(t)$
for $\mu=0.95$ (solid line) and $\mu=0.97$ (dashed line). Both systems have been perturbed at $t_0=10$ by a pulse of constant amplitude $p=1.7$ and duration $\Delta t = 0.4$. Note that, in agreement with the left panel, the system with the larger value of $\mu$ pulses before the one with the smaller value.}
\end{figure}

In order to understand the mechanism of the observed phenomenon,
we analyze the behavior of the master alone under the effect of a
single perturbation $I(t)=p_0\delta(t-t_0)$ acting at a certain
time $t_{0}$. The effect of this perturbation appears only as a
discontinuity of the, say, $x(t)$ variable at time $t_0$ as
$x(t_0^+)=x(t_0^-)+p_0$. The condition for the perturbation to be
larger than the excitability threshold, is that $x(t_0^+)>x_+$.
>From now on, we set the initial condition to be in the rest state,
$x(t_0^-)=x_-$, such the minimum value for the amplitude in order
to excite a pulse is $p_0 > 2\arccos \mu$ and the system develops
a pulse after a certain response time $t_{r}$. This time can be
precisely defined as the time it takes $x(t)$ to reach a given
reference value, e.g. $x_r = \pi/2$. From Eq. (2) we have $t_r =
\int_{x(t_0^+)}^{\pi/2} \frac{dx}{\mu -\cos x}$ which yields
\begin{equation}\label{eq:tr}
t_r=\frac{1}{\sqrt{1-\mu ^2}}\ln \left [\frac{(1-b)
(1+b^{-1}\tan \frac{x(t_0^+)}{2})}{(1+b)(1-b^{-1}\tan \frac{x(t_0^+)}{2})}\right ]
\end{equation}
where $b=\sqrt{\frac{1-\mu}{1+\mu}}$. In Fig. \ref{fig2} (left
panel) we plot the response time as a function of the parameter
$\mu$ for a given value of the perturbation amplitude $p_0$. Note
that below the excitability threshold, $p_0 < 2\arccos(\mu)$
(equivalently $\mu<\cos(p_0/2)$), $t_r$ does not exist. For
$\mu>\cos(p_0/2)$ the response time $t_r$ is a decreasing function
of $\mu$ which approaches zero as $\mu \rightarrow 1$. This result
shows that the response time of an Adler system to an
above-threshold external perturbation progressively decreases as
the Andronov bifurcation point ($\vert \mu \vert = 1$) is
approached, in agreement with the numerical result shown in Fig.
\ref{fig2} (right panel).

The fact that the response time decreases with lower excitability
threshold, and that in the coupled system the slave can emit
pulses that are not followed by a pulse in the master, suggest
that the mechanism for anticipation in the master-slave
configuration is that the slave has a lower excitability threshold
than the master. This is supported by the following qualitative
argument: Imagine that at $t=t_0$ both systems, master and slave,
are in the rest state $x(t_0^-)=y(t_0^-)=x_-$. The effect of the
perturbation changes both values to $x(t_0^+)=x(t_0^-)+p_0$,
$y(t_0^+)=y(t_0^-)+p_0$. Due to the coupling, the slave can be
considered to have at this time an effective
$\mu_{\textrm{eff}}(t_0)=\mu+K[x(t_0^+)-y(t_0^+-\tau)]=\mu+K p_0$.
Since $\mu_{\textrm{eff}}(t)>\mu$
also for all times $t$ such that $t_0\leq t < t_0+\tau $, 
the excitability threshold of the
slave has been reduced and the response time decreases.

\begin{figure}[h]
\begin{center}
\epsfig{file=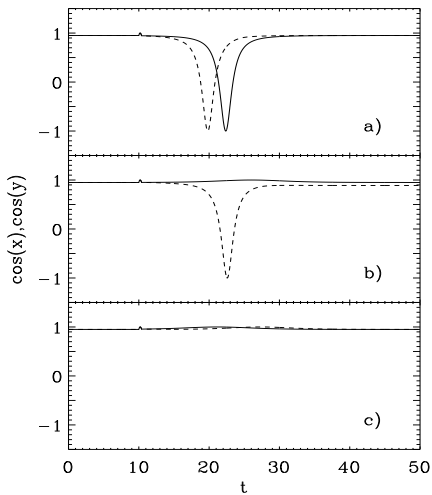,width=6cm,height=7cm}\end{center}
\caption{\label{fig3} Response of the master (solid line) and
slave (dashed line) for three different amplitudes of the singular
perturbation of duration $\Delta t=0.4$ at time $t_0=10$: (a) $p=1.7$, (b) $p=1.65$ and (c) $p=1.61$.
Other parameters are $\mu=0.95$, $\tau=5$ and $K=0.01$.}
\end{figure}

To give a more rigorous evidence for this explanation, we consider now 
two coupled
systems, Eqs.~(\ref{1}-\ref{2}), in the presence of a single 
perturbation which
we choose to be a pulse of constant amplitude $p$ and duration 
$\Delta t$
acting at time $t_0$ in which both systems are in the rest state 
$x(t_0^-)=y(t_0^-)=x_-$.
 The results are reported in fig \ref{fig3}.

\begin{figure}[h]
\begin{center}
\epsfig{file=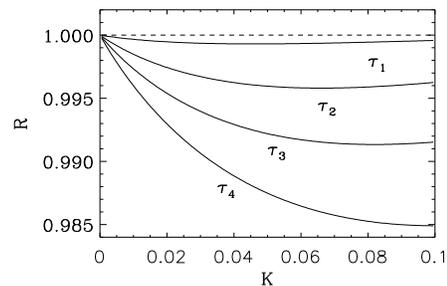,width=6cm,height=4cm}
\end{center}
\caption{\label{fig4} The ratio between the slave and the master
excitability threshold as a function of $K$ for $\tau _1=0.05$,
$\tau_2=0.2$, $\tau_3=0.35$ and $\tau_4=0.5$. Considered system have parameter
$\mu=0.95$. Perturbation is applied at time $t_0=10$ with magnitude $p=1.635$
and duration $\Delta t=0.4$. The dashed line corresponds
to the constant excitability threshold of the master.}
\end{figure}

For a sufficiently large perturbation, the master and the slave respond with an
excitable spike and the slave pulse anticipates the master pulse (fig. $3a$).
For small perturbation amplitude no pulses are generated and both  systems
respond proportionally to the applied stimulus (fig $4c$). However, an
intermediate amplitude of the perturbation triggers the emission of an
excitable pulse by the slave system while the master responds linearly (fig
$3b$). This confirms a lowering of the excitability threshold of the slave as
compared to the master, which is systematically found for all coupling
parameters that yield anticipating synchronization. In Fig.~\ref{fig4} we plot
the ratio $R$ between the minimum amplitude of the perturbation which generates
an excitable pulse in the slave and the minimum amplitude that generates a
pulse in the master. As shown in the figures, the effect of this particular
coupling scheme on the slave system is to lower its excitability threshold in
such way that the difference between the response time of the master and the
slave to an external perturbation equals approximately the delay in the
coupling term, $\tau$. It is worth noting that when $K$ or $\tau$ tend to
zero, not surprisingly the thresholds for the slave and the master tend to be
equal, while for large values of $\tau$ the difference between the two
thresholds is very large.

Clearly, the same reasoning can be followed if the perturbation applied to both
systems is a white noise process. This allows us to explain why the erroneous
synchronization events correspond to the slave system firing a pulse that is
not followed by a pulse in the master: for a particular noise level the master
response is proportional to the perturbation while the slave emits an excitable
pulse. By increasing the noise level both master and slave emit excitable
pulses, each pulse of the slave being anticipated respect to that of the
master.

\begin{figure}[h]
\begin{center}
\epsfig{file=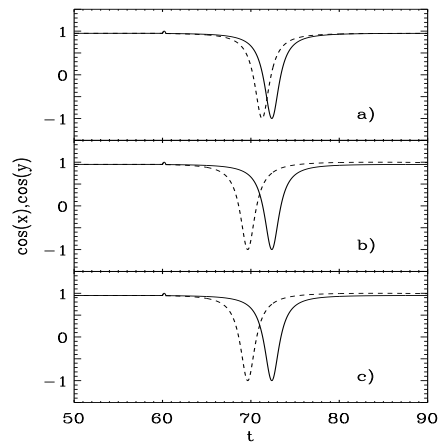,width=6cm,height=6cm}\end{center}
\caption{\label{fig5}Two coupled systems (master and slave) with a coupling
parameter $K=0.01$ and delay time (a) $\tau=1$, (b) $\tau=5$ and
(c) $\tau=50$. Both systems have $\mu=0.95$ and are perturbed at time $t_0=60$
with a pulse of magnitude $p=1.7$ and duration $\Delta t=0.4$.}
\end{figure}

Since, as we have shown, master and slave systems respond to
external perturbations with different response times, a question
which arises is whether it is possible to chose the parameters
such that the anticipation time is arbitrarily large, in
particular, larger than the master response time, $\tau
> t_{r}$, a result that would violate the causality
principle. In order to answer this question, we plot in Fig.
(\ref{fig5}) the results of integrating Eqs. \ref{1},\ref{2} under
the effects of a single perturbation for three
different values of the parameter $\tau$.  When $\tau < t_{r}$
(\ref{fig5}a) or $\tau \approx t_{r}$ (\ref{fig5}b), the
anticipation time is approximately equal to $\tau$. However, when
$\tau \gg t_{r}$, the anticipation time greatly differs from the
delay time, such that the slave anticipates the master by a time
interval always lower than $t_{r}$ (\ref{fig5}c). This is a
reasonable limit to the anticipation time: the pulse cannot
anticipate the perturbation  which created it. In other words,
master and slave are both "slaves" of the external perturbation,
although the presence of the master signal into the coupling term
contributes to lower the excitability threshold of the slave
leading to the anticipation phenomenon.

\begin{figure}[h]
\begin{center}
\epsfig{file=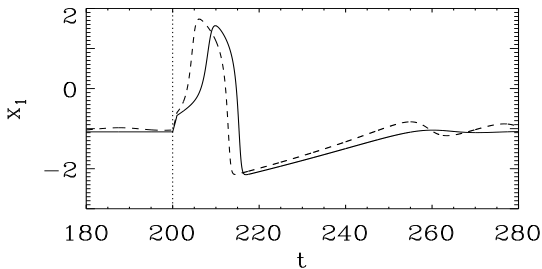,width=8cm,height=4cm}
\end{center}
\caption{\label{fig6} Time series for the variable $x_1$ of the FitzHugh-Nagumo system for $a=1.01$ (dashed line) and $a=1.08$ (solid line). In both cases it is $\epsilon=0.09$. As indicated by the vertical dotted line, the
system is perturbed at a time $t_0 = 200$ by a pulse of amplitude $p=0.4$ and duration  $\Delta t = 1$. Note that the response time decreases with increasing $a$.}
\end{figure}

In order to assess the generality of our hypothesis, we have also considered two
delayed coupled FitzHugh-Nagumo systems:
\begin{eqnarray}
(\dot{x_1}, \dot{x_2})  & = & (x_2 + x_1 - \frac{x_1^3}{3}, \epsilon (a-x_{1})) \label{FN:y1} \\
(\dot{y_1},\dot{y_2}) & = & (y_2 + y_1 - \frac{y_1^3}{3} + K (x_1 - y_1^{\tau}), \epsilon (a-y_1 ))\hspace{0.9cm}\label{FN:y2} 
\end{eqnarray}
In the excitable regime, which occurs when $\vert a \vert > 1$, the system possesses a single steady state. As the critical value $\vert a_c \vert = 1$ is approached, the excitability threshold is
lowered \cite{ste}. In this sense, the
control parameter $a$ plays the same role as the parameter $\mu$ in Adler's 
equation. In fact, we have
checked that also in this case the response time of the system to an external
perturbation decreases as the critical value $a_c$ is approached (see
Fig.~\ref{fig6}). 

\begin{figure}[h]
\begin{center}
\epsfig{file=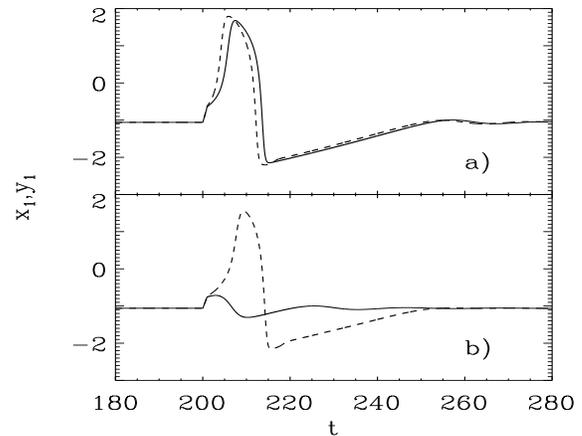,width=8cm,height=6cm}
\end{center}
\caption{\label{fig7} Response of the master ($x_1$, solid line) and
slave ($y_1$, dashed line) for two coupled FitzHugh-Nagumo systems  with $a=1.01$, $\epsilon=0.09$, $\tau=4$, $K=0.1$, after perturbation at $t_0=200$ by a pulse of amplitude $p$ and duration  $\Delta t = 1$. For large amplitude, $p=0.4$, case (a), both systems pulse whereas for the smaller amplitude, $p=0.3$, there is only pulse in the slave variable.}
\end{figure}

We now consider the unidirectionally delayed coupled system.
We find, as in the Adler's system, that the excitability threshold for
the slave is lower than that of the master, as shown in Fig.~\ref{fig7}
and that the maximum anticipation time is limited by the response time of the
master. Finally, we note that we have also found exactly the same phenomenology
for two delayed coupled Hodgkin-Huxley systems.

The ubiquity of this effect is, in our opinion, an indication that
the lowering of the excitability threshold of the slave in a
delayed coupling scheme is a general mechanism for anticipating
synchronization in excitable systems. This mechanism allows to
explain all the observations in the regime of anticipating
synchronization, in particular the erroneous firing of pulses in
the slave system. In addition, it evidences the causality of this
phenomenon: the master and slave systems follow the applied
external perturbations, although the response time of the slave
system is shorter due to the effects of the coupling. Moreover, we
have shown that the anticipation time is limited by the response
time of the master system. The relevance of this type of mechanism
for the synchronization of coupled chaotic systems is an open
question that will be studied in the near future.

We acknowledge financial support from MCYT (Spain) and FEDER through projects
TIC2002-04255-C04, BFM2000-1108 and BFM2001-0341-C02-01. S.B. acknowledges
financial support from MEC through sabbatical grant PR2002-0329.

\end{document}